# 基於假名憑證的健康照護系統
# The Pseudonymous Certificates for Healthcare Systems

**Abel C. H. Chen**

**Telecommunication Laboratories, Chunghwa Telecom Co., Ltd.**

## 摘要

**研究目的**：近年來隨著駭客技術的進步，資訊安全議題逐漸受到重視，特別是個資相關風險議題。為提供保密和可信任的資料傳輸環境，基於非對稱式密碼學的公開金鑰基礎建設可以提供私鑰簽章、公鑰驗章的機制。然而，在這個機制下，健康資料雖可以通過驗證來確認是來自特定用戶簽章的，但卻反而暴露了簽章者資訊，讓網路駭客可以對應出該健康資料所屬的用戶，對隱私產生威脅。有鑑於此，建立一套假名憑證(Pseudonymous Certificates, PCs)應用於健康照護系統保護用戶隱私是重要的研究議題。

**研究方法**：為建立一套基於假名憑證的健康照護系統，本研究主要修改蝴蝶金鑰擴展(Butterfly Key Expansion, BKE)方法，並且應用在健康照護系統。在系統中主要包含根憑證中心(Root Certificate Authority, RCA)、註冊憑證中心(Enrollment Certificate Authority, ECA)、假名憑證中心(Pseudonym Certificate Authority, PCA)、醫療院所登錄中心(Registration Authority, RA)、以及終端設備(End Entity, EE)(如：用戶端設備)。可由根憑證中心簽發憑證給註冊憑證中心、假名憑證中心、醫療院所登錄中心，使其成為系統中合法認證的實體。再由註冊憑證中心簽發設備憑證(類似設備的身份證)給終端設備(包含血壓計等)。後續健康照護患者使用終端設備(用戶端設備)量測生理資料時，可由註冊憑證中心確認是合法終端設備後，由假名憑證中心簽發多張假名憑證給終端設備(用戶端設備)，再由終端設備(用戶端設備)使用假名憑證來發送生理資料給終端設備(醫療院所)，可以保證資料完整性、不同否認性，同時避免身份資訊被盜取。

**結果與討論**：為驗證本研究提出的基於假名憑證的健康照護系統，採用美國國家標準技術局(National Institute of Standards and Technology, NIST)定義的安全強度來驗證系統的安全性。此外，由於蝴蝶金鑰擴展主要建構在橢圓曲線密碼學(Elliptic Curve Cryptography, ECC)的基礎上，本研究驗證在不同安全強度下的系統效率。其中，在安全強度 256 的情況下，金鑰擴展平均時間介於 24186.584 微秒和 57894.552 微秒之間，所以 1 秒內仍可以擴展 18 把以上的公鑰。

**結論**：本研究提出一套基於假名憑證的健康照護系統，在蝴蝶金鑰擴展方法可以提供安全且高效的假名憑證。通過假名憑證可以運用擴展後的私鑰進行簽章，並且通過擴展後的公鑰進行驗章，並無法從擴展後的公鑰推估原始公鑰，從而保證用戶隱私。

**關鍵詞**：假名憑證、蝴蝶金鑰擴展、健康照護系統

---

*通訊作者: Abel C. H. Chen (chchen.scholar@gmail.com)

# Abstract

**Objectives**: In recent years, with the advancement of hacker technology, information security issues have gradually been given more attention, especially those related to personal information risks. In order to provide a confidential and trustworthy data transmission environments, a public key infrastructure (PKI) based on asymmetric cryptography can provide private keys for signatures and public keys for verification. However, in this mechanism, although health data can be verified to confirm that it comes from a specific user signature, it actually exposes the information of the signer. Hackers on the internet may identify the user to whom the health data belongs for creating a threat to privacy. Therefore, it is an important research issue to establish a set of pseudonymous certificates (PCs) for use in healthcare systems to protect user privacy.

**Methods**: To establish a pseudonymous certificate-based healthcare system, this study mainly modifies the butterfly key expansion (BKE) mechanism and applies it to the healthcare system. The system mainly includes a Root Certificate Authority (RCA), an Enrollment Certificate Authority (ECA), a Pseudonym Certificate Authority (PCA), a Registration Authority (RA), and End Entities (EEs)(i.e. user devices). Certificates can be issued by the RCA to the ECA, PCA, and RA to make them legal entities in the system. The ECA then issues device certificates (similar to identification cards for devices) to the EEs (e.g. blood pressure monitors). When patients use EEs to measure physiological information, the RA verifies that the EE is legal based on the issued multiple pseudonym certificates by the PCA. The EE then uses the pseudonym certificates to send physiological information to the RA, ensuring data integrity and non-repudiation, while also preventing identity information from being stolen.

**Results and Discussions**: To verify the pseudonymous certificate-based healthcare system proposed in this study, the security of the system was verified using the security strengths defined by the National Institute of Standards and Technology (NIST) in the United States. Furthermore, as the BKE mechanism is primarily based on Elliptic Curve Cryptography (ECC), this study also verified the efficiency under different security strengths. Furthermore, under 256 of security strength, the mean computation time of key expansion is between 24186.584 microseconds and 57894.552 microseconds, so more than 18 public keys could be generated in one second.

**Conclusions**: This study proposes a pseudonymous certificate-based healthcare system in which the BKE mechanism provides secure and efficient pseudonymous certificates. By using the extended private key, signatures can be made, and verification can be done using the extended public key. It is not possible to infer the original public key from the extended public key for ensuring user privacy.

# 壹、前言

隨著資訊技術的快速發展和進步，駭客技術能力也不斷提升，資訊安全和隱私保護威脅問題層出不窮。特別是醫療領域與個人隱私密切相關，更是隱私保護的重要領域之一。例如，美國伊利諾州杜佩琪醫療集團(DuPage Medical Group)在2021年7月受到駭客攻擊，並且在事件中外洩60萬名病患的姓名、生日、醫療過程和日期、以及部分人士的社會安全碼等個人資訊(世界日報，2021)。如果駭客竊取這些個人資訊後，盜用這些個人資訊，將可能造成病患的財產損失或其他嚴重的影響。進行有鑑於此，建立一套醫療和健康照護資訊安全系統避免遭受駭客攻擊，保護病患的隱私是重要的研究議題。

為建置醫療和健康照護資訊安全系統，基於非對稱式密碼學的公開金鑰基礎建設(Public Key Infrastructure, PKI)是常用的架構，可以提供數位簽章和加解密服務(Höglund et al., 2020)。其中，目前採用的非對稱式密碼學主要包含RSA和橢圓曲線密碼學(Elliptic Curve Cryptography, ECC)，用來產製金鑰對(包含公鑰和私鑰)，並應用在各式各樣的醫療和健康照護服務(Kavitha, Alphonse, & Reddy, 2019)。在非對稱式密碼學和公開金鑰基礎建設可以提供數位簽章服務，保證簽發出來的訊息的可靠性，具有不可否認性，但卻也同時有暴露隱私的疑慮。因此，IEEE組織在近幾年提出蝴蝶金鑰擴展方法(Intelligent Transportation Systems Committee, 2022)，可以在既有的非對稱式密碼學和公開金鑰基礎建設上提供假名憑證，既保證簽發訊息是來自合法設備，也保障了用戶的隱私。

有鑑於蝴蝶金鑰擴展方法主要是針對車聯網領域設計(Intelligent Transportation Systems Committee, 2022)，本研究提出基於假名憑證的健康照護系統，設計適合醫療和健康照護領域的蝴蝶金鑰擴展方法和假名憑證機制。在系統中，可由根憑證中心(Root Certificate Authority, RCA)簽發憑證給註冊憑證中心(Enrollment Certificate Authority, ECA)、假名憑證中心(Pseudonym Certificate Authority, PCA)、醫療院所登錄中心(Registration Authority, RA)，使其成為系統中合法認證的實體。再由註冊憑證中心簽發設備憑證(類似設備的身份證)給終端設備(End Entity, EE)(包含血壓計等)。後續健康照護患者使用終端設備(用戶端設備)量測生理資料時，可由註冊憑證中心確認是合法終端設備後，由假名憑證中心簽發多張假名憑證給終端設備(用戶端設備)，再由終端設備(用戶端設備)使用假名憑證來發送生理資料給終端設備(醫療院所)，可以保證資料完整性、不同否認性，同時避免身份資訊被盜取。

本研究的主要貢獻條列如下：
1. 提出基於假名憑證的健康照護系統的系統架構，在公開金鑰基礎建設上建立非對稱金鑰技術加解密和數位簽章,保障終端設備(用戶端設備)量測到的生理資料可在密文情況下傳輸給終端設備(醫療院所)，並且僅有終端設備(醫療院所)可以解密。除此之外，生理資料是由終端設備(用戶端設備)進行簽章，確保生理資料的資料可靠性、不同否認性。
2. 提出基於假名憑證的健康照護系統的系統流程，本研究在蝴蝶金鑰擴展方法的基礎上設計假名憑證的系統流程，可以用來保障生理資料是由合法終端設備(用戶端設備)簽章，並且同時可以運用假名憑證來避免身份資訊被盜取。
3. 本研究採用美國國家標準技術局(National Institute of Standards and Technology,

NIST)定義的安全強度(Barker, 2020)來驗證基於假名憑證的健康照護系統在不同安全強度下的系統效率。

本文總共分為五個章節。本研究主要在蝴蝶金鑰擴展方法的基礎上設計基於假名憑證的健康照護系統，所以第貳節介紹橢圓曲線密碼學和蝴蝶金鑰擴展方法。第參節詳述本研究提出的基於假名憑證的健康照護系統，設計具備假名憑證的健康照護系統的系統架構，再設計結合蝴蝶金鑰擴展方法的系統流程。第肆節進行系統效能分析，驗證本研究提出的基於假名憑證的健康照護系統在不同安全強度下的系統效率。最後，第伍節總結本研究的主要貢獻，並討論未來的研究方向。

## 貳、研究背景

因為 IEEE 標準中的蝴蝶金鑰擴展方法主要建構在橢圓曲線密碼學的基礎上，所以本節將先介紹橢圓曲線密碼學，再介紹蝴蝶金鑰擴展方法。

### 一、橢圓曲線密碼學

為說明橢圓曲線密碼學，將在第貳.一.(一)節中定義橢圓曲線密碼學的相關函數和係數、第貳.一.(二)節介紹橢圓曲線數位簽章演算法(Elliptic Curve Digital Signature Algorithm, ECDSA)、第貳.一.(三)節介紹橢圓曲線迪菲-赫爾曼金鑰交換(Elliptic Curve Diffie–Hellman key exchange, ECDH)演算法、第貳.一.(四)節介紹橢圓曲線整合加密機制(Elliptic Curve Integrated Encryption Scheme, ECIES)(Aumasson, 2018)。

**(一) 橢圓曲線密碼學相關定義**

本節將說明橢圓曲線密碼學相關定義，分別從初始階段和產製金鑰階段來介紹橢圓曲線函數及其相關參數。

在初始階段，需先定義橢圓曲線函數(如：NIST 標準定義的 Weierstrass 函數)，如公式(1)所示。其中，將包含座標值$(x, y)$、係數$\alpha$、常數$\beta$、質數模數$n$、以及基點座標$G$。為建立安全的橢圓曲線函數，NIST 標準定義了多個橢圓曲線函數及其相關參數值，如：NIST P-256，在實作上可採用 NIST 標準(Chen et al., 2023)。

$$y^2 = x^3 + \alpha x^2 + \beta \mod n \tag{1}$$

在產製金鑰階段，先選擇合適的橢圓曲線函數及其相關參數(如：NIST P-256)，再產製隨機產生足夠大的整數作為私鑰$a$，然後運用$a$倍的基點座標$G$點之$x$座標值和作$y$座標值作為公鑰$A$ (即 $A = aG$)。其中，$a$倍的基點座標$G$點並非對$G$點之$x$座標值和作$y$座標值乘上$a$，而是通過橢圓曲線點的加法計算方法和雙倍計算方法計算取得(Chen et al., 2023)，所以在產製公鑰的過程需較多的計算時間。

**(二) 橢圓曲線數位簽章演算法**

本節將說明橢圓曲線數位簽章演算法，假設存在橢圓曲線金鑰對(私鑰為$a$、公鑰為$A = aG$)，運用私鑰$a$對訊息$m$進行簽章，並且可以用公鑰$A$對簽章內容進行驗章(Johnson & Menezes, 1999)。

在簽章階段，產生隨機數$k$，計算$K = kG$，取得$K$座標$(x_K, y_K)$，再運用公式(2)計算$s$，取得簽章$(K, s)$(Johnson & Menezes, 1999)。

$s = [(h + px_K) / k](\mod n)$     (2)

在驗章階段，運用公式(2)~公式(6)計算橢圓曲線點 $D$，並且當 $D = kG = K$ 時，代表驗章通過(公式證明如公式(7))(Johnson & Menezes, 1999)。

$w = 1 / s = [k / (h + px_K)](\mod n)$     (3)

$u = hw = h / s = hk / (h + px_K)$     (4)

$v = x_K w = x_K / s = x_K k / (h + px_K)$     (5)

$D = uG + vP = uG + vpG = (u + vp)G$     (6)

$u + vp = [hk / (h + px_K)] + [px_K k / (h + px_K)] = k$     (7)

### (三) 橢圓曲線迪菲-赫爾曼金鑰交換

本節將說明橢圓曲線迪菲-赫爾曼金鑰交換演算法，假設使用者 1 有橢圓曲線金鑰對(私鑰為 $q$、公鑰為 $Q = qG$)、使用者 2 有橢圓曲線金鑰對(私鑰為 $c$、公鑰為 $C = cG$)，並且由這兩組橢圓曲線金鑰對進行金鑰協商，取得共享金鑰 $x_M$ (Subramanian & Tamilselvan, 2020)。

在金鑰協商階段，使用者 1 傳其公鑰 $Q$ 給使用者 2，使用者 2 可搭配其私鑰 $c$ 計算 $cQ$；使用者 2 傳其公鑰 $C$ 給使用者 1，使用者 1 可搭配其私鑰 $q$ 計算 $qC$。並且，由公式(8)可以證明橢圓曲線點 $cQ$ 和橢圓曲線點 $qC$ 為相同的橢圓曲線點 $M$，取得橢圓曲線點 $M$ 座標$(x_M, y_M)$的 $x$ 座標 $x_M$ 作為共享金鑰(Subramanian & Tamilselvan, 2020)。

$cQ = cqG = qcG = qC = M$     (8)

### (四) 橢圓曲線整合加密機制

本節將說明橢圓曲線整合加密機制，假設使用者 1 有橢圓曲線金鑰對(私鑰為 $q$、公鑰為 $Q = qG$)、使用者 2 有橢圓曲線金鑰對(私鑰為 $c$、公鑰為 $C = cG$)，並且使用者 2 對訊息 $m$ 進行加密為密文 $m'$，再由使用者 1 對密文 $m'$ 進行解密為明文 $m$ (Martínez, Encinas, & Ávila, 2010)。

在金鑰協商階段，運用橢圓曲線迪菲-赫爾曼金鑰交換演算法可取得共享金鑰 $x_M$，如第貳.一.(三)節所述。其中，以 $x$ 座標 $x_M$ 的前 $l$ 個 Bytes 作為訊息鑑別碼(Message Authentication Code, MAC)演算法簽章金鑰$\phi$，再 $x$ 座標 $x_M$ 的第 $l$ 個後的 Bytes 作為進階加密標準(Advanced Encryption Standard, AES)演算法加解密金鑰$\varphi$。其中，$l$ 的取值，主要根據進階加密標準演算法的金鑰長度來決定。後續使用者雙方以金鑰$\phi$進行簽章，再以金鑰$\varphi$進行加解密(Martínez, Encinas, & Ávila, 2010)。

在加密階段，使用者 2 執行金鑰協商，取得金鑰$\varphi$以進階加密標準演算法對訊息 $m$ 進行加密為密文 $m'$ (Martínez, Encinas, & Ávila, 2010)。

在加密階段，使用者 1 執行金鑰協商，取得金鑰$\varphi$以進階加密標準演算法對密文 $m'$ 進行解密為明文 $m$ (Martínez, Encinas, & Ávila, 2010)。

## 二、蝴蝶金鑰擴展方法

蝴蝶金鑰擴展方法的主要流程包含終端設備產製毛蟲金鑰對，再由登錄中心基於毛蟲公鑰產製繭公鑰，並且由假名憑證中心基於繭公鑰產製蝴蝶公鑰，最後由終端設備產製繭私鑰和蝴蝶私鑰(Intelligent Transportation Systems Committee, 2022)，如圖 1 所示。

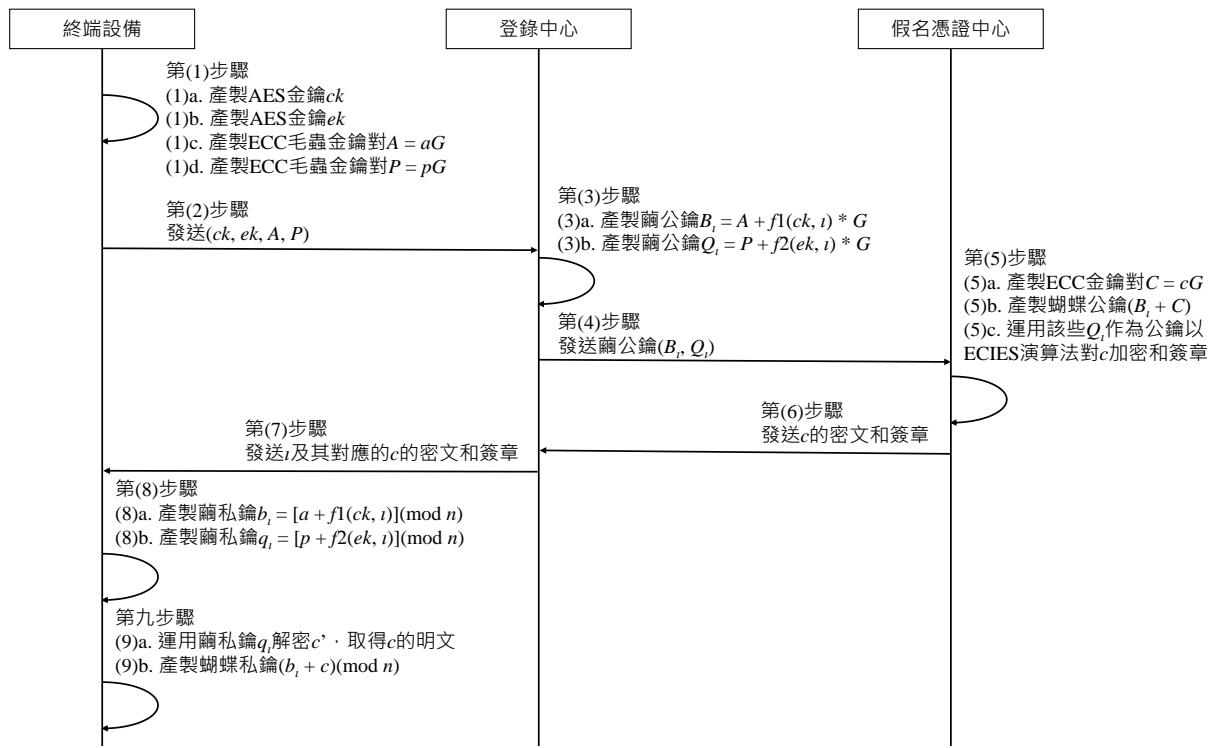

圖 1、蝴蝶金鑰擴展方法(Intelligent Transportation Systems Committee, 2022)

## (一) 具體方法流程

具體方法流程(Intelligent Transportation Systems Committee, 2022)說明如下：

(1). 由終端設備產製 AES 金鑰及 ECC 金鑰對，具體方法流程包含有：
   a. 產製 AES 金鑰 $ck$，作為簽章使用。其中，參數 $ck$ 是對稱式金鑰。
   b. 產製 AES 金鑰 $ek$，作為加密使用。其中，參數 $ek$ 是對稱式金鑰。
   c. 產製 ECC 金鑰對 $A = aG$，作為毛蟲金鑰對，簽章使用。其中，參數 $a$ 是私鑰、參數 $A$ 是公鑰、參數 $G$ 是橢圓曲線的基準點。
   d. 產製 ECC 金鑰對 $P = pG$，作為毛蟲金鑰對，簽章使用。其中，參數 $p$ 是私鑰、參數 $P$ 是公鑰、參數 $G$ 是橢圓曲線的基準點。

(2). 由終端設備將產製的對稱式金鑰及毛蟲公鑰($ck, ek, A, P$)發送給登錄中心。

(3). 由登錄中心基於毛蟲公鑰產製繭公鑰，具體方法流程包含有：
   a. 產製繭公鑰 $B_l = A + f1(ck, \iota) * G$。其中，參數 $\iota$ 是增量整數；函數 $f1$ 是基於 AES 加密演算法的擴展函數，可運用 AES 金鑰 $ck$ 加密參數 $\iota$ 值，得到整數密文，避免被攻擊者取得參數 $\iota$ 明文的情況下從繭公鑰推導出毛蟲公鑰。
   b. 產製繭公鑰 $Q_l = P + f2(ek, \iota) * G$。其中，參數 $\iota$ 是增量整數；函數 $f2$ 是基於 AES 加密演算法的擴展函數，可運用 AES 金鑰 $ek$ 加密參數 $\iota$ 值，得到整數密文，避免被攻擊者取得參數 $\iota$ 明文的情況下從繭公鑰推導出毛蟲公鑰。

(4). 登錄中心將產製的繭公鑰($B_l, Q_l$)發送給假名憑證中心。

(5). 假名憑證中心基於繭公鑰產製蝴蝶公鑰，蝴蝶公鑰可以作為假名憑證的公鑰加密使用，具體方法流程包含有：

a. 產製 ECC 金鑰對 $C = cG$。其中，參數 $c$ 是私鑰、參數 $C$ 是公鑰、參數 $G$ 是橢圓曲線的基準點。
   b. 產製蝴蝶公鑰$(B_\iota + C)$。
   c. 運用 $Q_\iota$ 作為公鑰以橢圓曲線整合加密機制對 $c$ 加密和簽章，其中 $c$ 的密文係 $c'$。
(6). 假名憑證中心發送密文 $c'$ 和簽章給登錄中心。
(7). 登錄中心發送 $\iota$ 值及其對應的密文 $c'$ 和簽章給終端設備。
(8). 終端設備根據 $\iota$ 值產製繭私鑰，具體方法流程包含有：
   a. 產製繭私鑰 $b_\iota = [a + f1(ck, \iota)] \pmod n$。其中，函數 $f1$ 是基於 AES 加密演算法的擴展函數，可運用 AES 金鑰 $ck$ 加密參數 $\iota$ 值，得到整數密文、參數 $n$ 是橢圓曲線的階。
   b. 產製繭私鑰 $q_\iota = [p + f2(ek, \iota)] \pmod n$。其中，函數 $f2$ 是基於 AES 加密演算法的擴展函數，可運用 AES 金鑰 $ek$ 加密參數 $\iota$ 值，得到整數密文、參數 $n$ 是橢圓曲線的階。
(9). 終端設備使用繭私鑰 $q_\iota$ 解密取得 $c$ 值，並且運用繭私鑰 $b_\iota$ 和 $c$ 值產製蝴蝶私鑰，蝴蝶私鑰可作為假名憑證的私鑰簽章或解密使用，具體方法流程包含有：
   a. 運用繭私鑰 $q_\iota$ 解密密文 $c'$，取得明文 $c$。
   b. 產製蝴蝶私鑰$(b_\iota + c) \pmod n$。其中，參數 $n$ 是橢圓曲線的階。

**(二) 安全性和隱私討論**

蝴蝶金鑰擴展方法可以保障終端設備的隱私，其理論基礎說明如下：

‧ 對登錄中心而言，因為登錄中心不知道明文 $c$ 值和 $C$ 值，所以無法從蝴蝶公鑰推導對應的繭公鑰。此外，登錄中心亦不知毛蟲私鑰，所以也無法推導繭私鑰和蝴蝶私鑰(Intelligent Transportation Systems Committee, 2022)。

‧ 對假名憑證中心而言，因為假名憑證中心不知道擴展函數加密後的 $\iota$ 值，所以無法從繭公鑰推導對應的毛蟲公鑰。此外，假名憑證中心亦不知毛蟲私鑰，所以也無法推導繭私鑰和蝴蝶私鑰(Intelligent Transportation Systems Committee, 2022)。

## 參、基於假名憑證的健康照護系統

本節詳述本研究提出的基於假名憑證的健康照護系統,在第參.一節描述系統架構、第參.二節描述系統流程，分述如下。

### 一、系統架構

基於假名憑證的健康照護系統主要包含根憑證中心、註冊憑證中心、假名憑證中心、醫療院所登錄中心、終端設備(用戶端設備)、以及終端設備(醫療院所)(如圖 2 所示)，以下對每個元件進行說明。

(1). 根憑證中心：根憑證中心是公開金鑰基礎建設中所有設備信任的源頭，所有設備都信任根憑證中心簽發出來的憑證，及其下屬的憑證鏈。由此建構應用在健康照護系

統的公開金鑰基礎建設。
(2). 註冊憑證中心：註冊憑證中心可以為每一台終端設備出廠時，由註冊憑證中心簽發終端設備的初始憑證。在本節中，終端設備(用戶端設備)初始金鑰對是$(a, A)$，並且 $a$ 為私鑰、$A$ 為公鑰，由註冊憑證中心簽發的終端設備(用戶端設備)憑證中具有公鑰 $A$ 資訊；終端設備(醫療院所)初始金鑰對是$(h, H)$，並且 $h$ 為私鑰、$H$ 為公鑰，由註冊憑證中心簽發的終端設備(醫療院所)憑證中具有公鑰 $H$ 資訊。
(3). 假名憑證中心：假名憑證中心可以執行蝴蝶金鑰擴展方法，為終端設備(用戶端設備)產製蝴蝶公鑰(即假名憑證)。
(4). 醫療院所登錄中心：醫療院所登錄中心可以根據終端設備(用戶端設備)的憑證(即毛蟲公鑰)向註冊憑證中心確認終端設備(用戶端設備)的合法性。終端設備(用戶端設備)確屬合法設備，再執行蝴蝶金鑰擴展方法，產製繭公鑰，並向假名憑證中心請求產製蝴蝶公鑰(即假名憑證)。
(5). 終端設備(用戶端設備)：終端設備(用戶端設備)是病患自家使用的生理資料量測設備，可以量測病患的生理資料後發送給終端設備(醫療院所)。並且為了保護隱私，終端設備(用戶端設備)可以執行蝴蝶金鑰擴展方法取得假名憑證後，以擴展後私鑰進行簽章。
(6). 終端設備(醫療院所)：終端設備(醫療院所)是醫師端使用的設備，可以取得病患的生理資料，並且終端設備(醫療院所)可以驗證生理資料是來自合法的終端設備(用戶端設備)。除此之外，終端設備(醫療院所)會與終端設備(用戶端設備)協商擴展值 $t$，在傳輸過程中會以終端設備(醫療院所)擴展後公鑰加密，保證每個終端設備(用戶端設備)都是以不同公鑰加密，並且終端設備(醫療院所)擴展後私鑰解密取得明文。

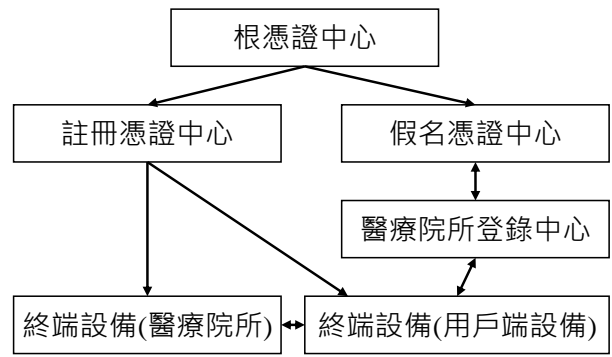

圖 2、基於假名憑證的健康照護系統的系統架構圖

## 二、系統流程

本節說明基於假名憑證的健康照護系統的系統流程(如圖 3 所示)，詳述如下。

**第(1)步驟**：初始階段，終端設備(用戶端設備)和終端設備(醫療院所)出廠時，由註冊憑證中心簽發終端設備的初始憑證。終端設備(用戶端設備)的初始金鑰對是$(a, A)$，終端設備(醫療院所)的初始金鑰對是$(h, H)$。

**第(2)步驟**：由終端設備(用戶端設備)、醫療院所登錄中心、以及假名憑證中心三方執行蝴蝶金鑰擴展方法(如圖 1 所示)，可以產生終端設備(用戶端設備)的假名憑證，並

且網路上的其他設備無法取得假名憑證與原始終端設備的對應關係。

**第(3)步驟**：終端設備(用戶端設備)取得假名憑證金鑰對$(s, S)$，並且$s$為擴展後私鑰、$S$為擴展後公鑰。

**第(4)步驟**：終端設備(醫療院所)與終端設備(用戶端設備)協商擴展值$t$。例如：病患就醫時由醫師當面提供一個序號 $t$ (由終端設備(醫療院所)隨機產生或病患在醫院的序號)，由病患回家後輸入序號$t$到終端設備(用戶端設備)。

**第(5)步驟**：終端設備(用戶端設備)可以根據序號 $t$ 搭配終端設備(醫療院所)公鑰 $H$ 運用公式(9)計算擴展後公鑰$Z$。終端設備(醫療院所)可以根據序號 $t$ 搭配終端設備(醫療院所)私鑰 $h$ 運用公式(10)計算擴展後私鑰$z$。

$$Z = zG = tG + H \tag{9}$$
$$z = t + h \tag{10}$$

**第(6)步驟**：終端設備(用戶端設備)可以用其擴展後的金鑰對$(s, S)$和終端設備(醫療院所)其擴展後的金鑰對$(z, Z)$，運用橢圓曲線整合加密機制加密生理資料。

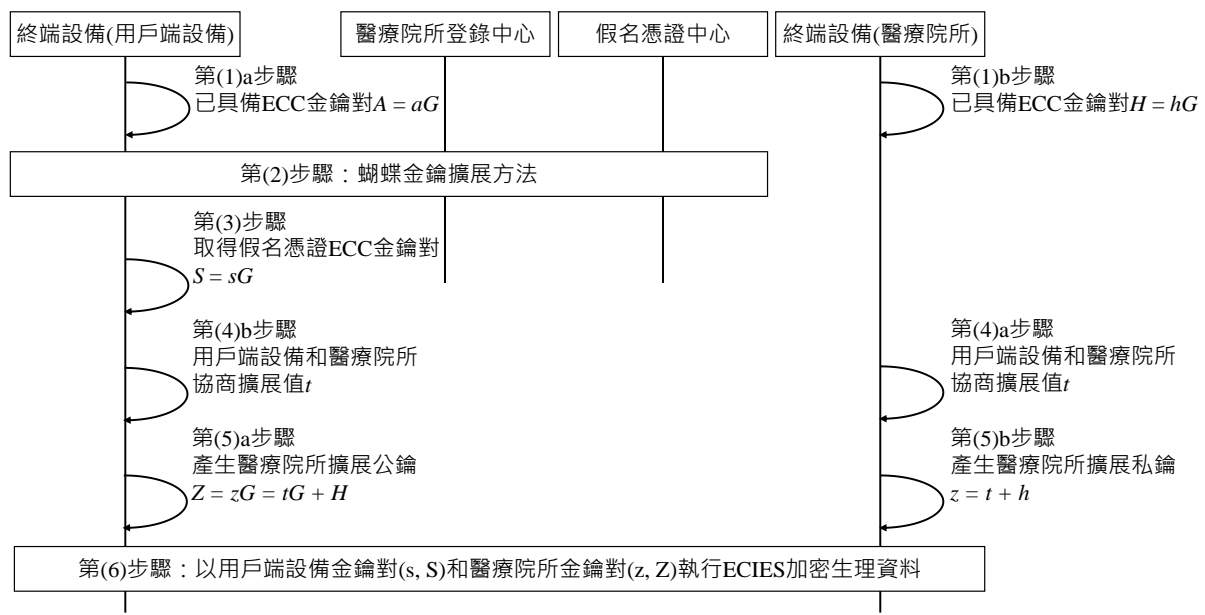

圖3、基於假名憑證的健康照護系統的系統流程圖

## 肆、實驗結果與討論

為驗證本研究提出的基於假名憑證的健康照護系統，採用美國國家標準暨技術研究院所規範的安全強度標準(Barker, 2020)在實際環境來驗證不同方法做金鑰擴展的效率。可將安全強度分為五個等級，並每個等級對應密碼學的要求來實作。除此之外，本研究總共設計 4 個實驗情境，分別在每個實驗中各執行 1000 次進行比較，詳細說明如下。

- 實驗1：從毛蟲公鑰擴展成繭公鑰，並只擴展成1個繭公鑰。
- 實驗2：從繭公鑰擴展成蝴蝶公鑰，並只擴展成1個蝴蝶公鑰。
- 實驗3：從毛蟲公鑰擴展成繭公鑰，並擴展成20個繭公鑰。
- 實驗4：從繭公鑰擴展成蝴蝶公鑰，並擴展成20個蝴蝶公鑰。

由實驗結果如表 1 所示，在安全強度 80 時，實驗 1 花費的金鑰擴展平均時間為

55589.152 微秒,標準差為 24823.229。在實驗中,金鑰擴展平均時間介於 18172.760 微秒和 54905.265 微秒之間,所以 1 秒內可以擴展 18 把以上的公鑰。

表 1、金鑰擴展時間(單位:微秒)

| 安全強度 | 實驗 1 | 實驗 2 | 實驗 3 | 實驗 4 |
|---|---|---|---|---|
| 80 | 55589.152 (24823.229) | 18172.760 (5672.441) | 37656.958 (4003.968) | 18684.688 (4945.681) |
| 112 | 54905.265 (9530.397) | 32261.409 (14346.036) | 38377.835 (4676.397) | 25290.976 (12080.073) |
| 128 | 54347.954 (9686.334) | 21711.568 (2651.270) | 46556.214 (5849.735) | 21381.615 (9460.058) |
| 192 | 59667.727 (10590.645) | 19833.719 (2378.780) | 53909.124 (11794.927) | 23523.532 (6527.510) |
| 256 | 57025.756 (11479.287) | 24186.584 (5284.956) | 57894.552 (10633.871) | 28129.762 (5538.829) |

由於從毛蟲公鑰擴展成繭公鑰時,蝴蝶金鑰擴展方法需計算兩個繭公鑰,並且這兩個繭公鑰都需要計算 AES 演算法基礎的擴展函數對擴展值加密,所以需要花費大量的運算時間在擴展值加密及其累加擴展值密文倍數的橢圓曲線基準點 $G$;因此,在實驗 1 和實驗 3 花費的金鑰擴展時間較長。除此之外,從繭公鑰擴展成蝴蝶公鑰時,蝴蝶金鑰擴展方法需產製一組 ECC 金鑰對 $C = cG$,所以需要花費大量的運算時間在計算 $c$ 倍數的橢圓曲線基準點 $G$。不過由於在計算蝴蝶公鑰時,蝴蝶金鑰擴展方法需產製一組 ECC 金鑰對,所以實驗 2 和實驗 4 金鑰擴展時間較實驗 1 和實驗 3 時短。

## 伍、結論與未來研究

本研究建立一套基於假名憑證的健康照護系統,修改蝴蝶金鑰擴展方法,並且應用在健康照護系統。在系統中主要包含根憑證中心、註冊憑證中心、假名憑證中心、醫療院所登錄中心、以及終端設備(用戶端設備)和終端設備(醫療院所)。基於假名憑證的健康照護系統在公開金鑰基礎建設上建立非對稱金鑰技術加解密和數位簽章,保障終端設備(用戶端設備)量測到的生理資料可在密文情況下傳輸給終端設備(醫療院所),並且僅有終端設備(醫療院所)可以解密。除此之外,生理資料是由終端設備(用戶端設備)進行簽章,確保生理資料的資料可靠性、不同否認性。以及,本研究在蝴蝶金鑰擴展方法的基礎上設計假名憑證的系統流程,可以用來保障生理資料是由合法終端設備(用戶端設備)簽章,並且同時可以運用假名憑證來避免身份資訊被盜取。

由於目前的蝴蝶金鑰擴展在毛蟲公鑰擴展成繭公鑰時,需要花費大量的計算時間,在未來可以考慮設計更快的金鑰擴展方法。除此之外,雖然目前橢圓曲線密碼學是主流的非對稱式金鑰密碼學的方法,但已經被證明可能遭受量子計算攻擊,所以未來可以設計後量子密碼學方法來提升安全性。

# 參考文獻